\documentclass[a4paper,10pt]{article}
\usepackage{amsfonts}
\usepackage{amsmath}
\usepackage{amssymb}
\usepackage[dvips]{graphicx}

\DeclareGraphicsExtensions{.bmp,.png,.pdf,.jpg,.eps}

\begin{document}

\title{Effect of an external electric field on local magnetic moments in
silicene}
\author{J. Villarreal$^{\dag }$, F. Escudero$^{\dag }$, J. S. Ardenghi$%
^{\dag }$\thanks{%
email:\ jsardenghi@gmail.com, fax number:\ +54-291-4595142} and P. Jasen$%
^{\dag }$ \\
%EndAName
IFISUR, Departamento de F\'{\i}sica (UNS-CONICET)\\
Avenida Alem 1253, Bah\'{\i}a Blanca, Buenos Aires, Argentina}
\maketitle

\begin{abstract}
In this work we analyze the effects of the application of an external
electric field in the formation of a local magnetic moment in silicene. By
adding an impurity in a top site in the host lattice and computing the real
and imaginary part of the self-energy of the impurity energy level, the
polarized density of states is used in order to obtain the occupation number
of the up and down spin formation in the impurity considering the mean field
approximation. Unequal occupation numbers is the precursor of a formation of
a local magnetic moment and this depends critically on the Hubbard
parameter, the on-site energy of the impurity, the spin-orbit interaction in
silicene and the electric field applied. In particular, it is shown that in
the absence of electric field, the boundary between the magnetic and
non-magnetic phases increases with the spin-orbit interaction with respect
to graphene with a top site impurity and shrinks and narrows it when the
electric field is turned on. The electric field effect is studied for
negative and positive on-site impurity energies generalizing the results
obtained in the literature for graphene.
\end{abstract}

\section{Introduction}

In solid state physics, two dimensional (2D) systems have become one of the
most significant topics, where applications in nanoelectronics and
spintronics become possible due to the exotic electronic structures of these
2D materials(\cite{guz}, \cite{hou}). The most well known is graphene, a two
dimensional honeycomb lattice of carbon atoms \cite{geim}, but other new
developed 2D materials have arisen, such as molybdenum disulfide (MoS$_{2}$) 
\cite{chu}, silicene \cite{spen}, germanene (\cite{gill} and \cite{bian}),
phosphorene \cite{yar11}, transition-metal dichalcogenides and hexagonal
boron nitride \cite{roldan} and two-dimensional SiC monolayers \cite{yar18},
that are similar to graphene but with different atoms at each lattice site
but with a buckled structure. Due to the intrinsic low electron and phonon
densities, 2D materials are ideal platforms to host single atoms with
magnetic moments pointing out-of-plane with potential applications in
current information nanotechnology. Among these materials, silicene is
particularly interesting thanks to its compatibility with the current
Si-based electronic technology. In 2010, the synthesis of silicene
nanoribbons on the anisotropic Ag(110) surface and on Ag(111) was reported (%
\cite{pad}, \cite{vogt}), showing that silicene has a larger in-plane
lattice constant, with two interpenetrating sublattices displaced vertically
with respect to each other due to the $sp^{3}$ hybridization. In turn, the
buckling of silicene can be influenced by the interaction with a ZrB$_{2}$
substrate which allows to tune the band gap at the $K$ or $K^{\prime }$
points in the Brillouin zone \cite{fleu}. By applying the tight binding
model on silicene it is possible to compute the long wavelength
approximation in order to obtain an effective Dirac-like Hamiltonian (\cite%
{guz} and \cite{hou2} and \cite{wink}), and around the Fermi energy, the
charge carriers behaves as massive Dirac fermions in the $\pi $ bands moving
with a Fermi velocity $v_{F}=5.5\times 10^{5}$ m/s (\cite{dz} \cite{lebe}).\
The layer separation between the sublattices in silicene due to its buckled
structure, is suitable for application of external fields in order to open a
bandgap that introduces diagonal terms in the Hamiltonian written in the $A$
and $B$ sublattices (\cite{drum}, \cite{ni}, \cite{spen}).The spin-orbit
interaction (SOI) in silicene is about $3.9$ meV, larger than that of
graphene, where is of the order of $10^{-3}$ meV (\cite{liu}, \cite{yao}).
The large SOI allows the quantum spin Hall effect to be observed which
implies that silicene becomes a topological insulator (\cite{liu2}, \cite%
{eza}). The interplay between the SOI and external electric field can induce
transitions from topological to band insulators allowing valley effects in
the conductivity (\cite{drum}, \cite{yar4}).

When impurity atoms are deposited on graphene or silicene they can be
adsorbed on different adsorption sites, where the most usual is the six-fold
hollow site of the honeycomb lattice, on top of a carbon or silicon atom or
the two-fold bridge site of neighboring atoms of the host lattice (\cite%
{roten}, \cite{jsa1}, \cite{jsa2}, \cite{jsa3} and \cite{jsa4}). In turn,
adatoms bonded to the surface of graphene can lead to a quasi-localized
state where the wave function includes contributions from the orbitals of
neighboring carbon atoms (\cite{skr}, \cite{sofo}).

In particular, when the impurity atoms are magnetic, the strong coupling
between the localized magnetic state of the adatom and the band of the 2D
host lattice allows non-trivial effects in the static properties of the
system, such as the Kondo effect, where the local density of states show a
resonance at the Fermi level \cite{hew} due to the screening of the magnetic
moment of the adatom by the sourrounding itinerant electrons. In graphene,
the Kondo effect has been reported with Co adatoms spectroscopy \cite{Ren},
but not much is known about the spectral features or Kondo effect in other
2D materials such as silicene. In silicene, the effect of different magnetic
adatoms has been studied by using density functional theory (\cite{A}, \cite%
{B}, \cite{C} and \cite{D}), showing that silicene is able to form strong
bonds with transition metals due to its buckled form and that its properties
can be tailored to design batteries \cite{liuu}.

In turn, it has been shown that the magnetic properties of 2D materials are
very sensitive to the SOI and the application of external electric and
magnetic fields (\cite{fede1}, \cite{fede2}). These properties can be
altered when impurity atoms are introduced in the material because it
induces the formation of local magnetic moments. Thus, while there are
several numerical studies about transition metal adsorption in silicene and
other two-dimensional materials, not much is known about the dependence of
the localized magnetic moment on the applied external electric field. The
tight-binding method combined with the mean-field approximation \cite{ander}
allows studying the dependence of the local magnetism with the strong
correlation effects of the inner shell electrons, parametrized by the
on-site Hubbard contribution and the hybridization of the impurity orbital
with the host lattice. By computing the spin-polarized density of states
with Green function methods it is possible to obtain the occupation number
of each spin in the adatom (\cite{jsa1}, \cite{jsasili}). Moreover, the
effect of the SOI and an external electric field can tailor the magnetic
properties due to the interplay of the level broadening and the sublattice
asymmetry that induces a bandgap.

Motivated by this, in this paper we study the magnetic regime of the
impurity atom as a function of the Fermi level, the Hubbard parameter, the
SOI and an external electric field and we compare it with those obtained in
graphene. In particular we will consider impurity atoms adsorbed in a top
site in silicene with on-site energy below and above the Dirac point. Based
on these results it is possible to study the formation of localized magnetic
states in the impurities and their dependence with the external electric
field and the asymmetric hybridization with the host lattice. In turn, the
boundary between different magnetic phases can be approximated in terms of
the Fermi energy and\ the Hubbard parameter. In this sense, while there are
several works with ab-initio calculations with different transition metals,
not much is known about the magnetic features of silicene with respect to
the different parameters in the Hamiltonian. This work will be organized as
follows: In section II, the tight-binding model with adatoms is introduced
and the Anderson model in the mean-field approximation is applied to
silicene. In section III, the results are shown, and a discussion is given
and the principal findings of this paper are highlighted in the conclusion.

\section{Theoretical model}

The tight-binding Hamiltonian of silicene with spin-orbit coupling and a
perpendicular electric field reads (see \cite{spen})%
\begin{equation}
H_{0}=-t\underset{\left\langle i,j\right\rangle ,s}{\overset{}{\sum }}%
a_{i,s}^{\dag }b_{j,s}+h.c.+\frac{i\lambda _{SO}}{3\sqrt{3}}\underset{%
\left\langle \left\langle i,j\right\rangle \right\rangle ,s}{\overset{}{\sum 
}}s(\nu _{ij}a_{i,s}^{\dag }a_{j,s}+\nu _{ij}b_{i,s}^{\dag }b_{j,s})-elE_{z}%
\underset{i,s}{\overset{}{\sum }}\mu _{i}(a_{i,s}^{\dag
}a_{j,s}+b_{i,s}^{\dag }b_{j,s})  \label{s1}
\end{equation}%
where $a_{i,s}^{\dag }$($a_{j,s}$) are the creation (annihilation) operators
in the sublattice $A$ and $b_{i,s}^{\dag }$($b_{j,s}$) are the creation
(annihilation) operators in the sublattice $B$ of silicene in the site $i$
with spin~$s=\pm 1$. The first term of the last equation is the kinetic
energy which is $t_{gr}=2.7$eV for graphene and $t_{sil}=1.6$eV for
silicene. The second term represents the effective spin-orbit coupling with $%
\lambda _{SO}=3.9$ meV for silicene (see \cite{spen}) and $\nu _{ij}=(%
\mathbf{d}_{i}\mathbf{\times d}_{j})/\left\vert \mathbf{d}_{i}\mathbf{\times
d}_{j}\right\vert =\pm 1$, depending on the orientation of the two nearest
neighbor bonds $\mathbf{d}_{i}$ and $\mathbf{d}_{j}$ that connect the next
nearest neighbors $\mathbf{d}_{ij}$ (see \cite{manuel} and \cite{kane}). The
last term is the staggered sublattice potential with $\mu _{i}=+1(-1)$ for
the $A$ and $B$ sublattice sites, where the buckling for silicene is $%
l_{sil}=0.23$\AA\ \cite{spen} and $e$ is the electron charge. We are not
considering the Rashba spin-orbit coupling because it has a negligible
effect on the dispersion relation, being comparable to $\lambda _{so}$ only
at the near edge of the Brillouin zone \cite{zare}.

\begin{figure}[tbp]
\centering\includegraphics[width=90mm,height=90mm]{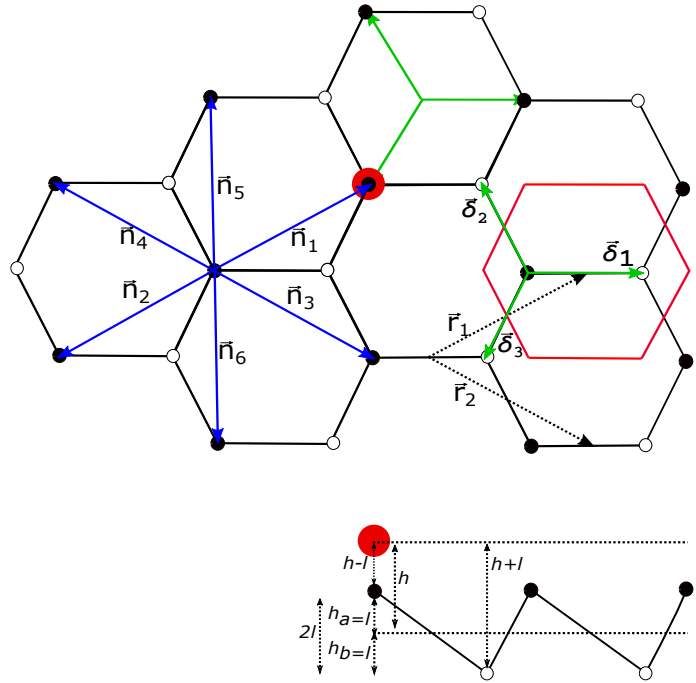}
\caption{Up. Silicene honeycomb lattice (black and white dots are silicon
atoms). Green arrows represent nearest neighbors and blue arrows represent
next-nearest neighbors, $\mathbf{r}_{1}$ and $\mathbf{r}_{2}$ are the
lattice vectors and the red hexagon is a particular Bravais lattice. Red
point represent impurity adsorbed on a top site. Down: Side view of silicene
with the adsorbed impurity where $l(h)$ is the distance of each
sublattice(impurity) with respect the middle of the buckling.}
\label{siligeom}
\end{figure}

The basis vectors for the hexagonal Bravais lattice can be written as $%
\mathbf{R}_{n,m}=n\mathbf{a}_{1}+m\mathbf{a}_{2}$ where $n,m$ are integer
numbers, $\mathbf{a}_{1}=\frac{a}{2}(3,\sqrt{3},0)$ and $\mathbf{a}_{2}=%
\frac{a}{2}(3,-\sqrt{3},0)$ are the primitive basis vectors (see red
hexagonal in figure \ref{siligeom}). Considering the Fourier transform of
the creation and annihilation operators $a_{j,s}=\frac{1}{\sqrt{N}}\overset{}%
{\underset{\mathbf{k}}{\sum }}e^{-i\mathbf{kR_{j}}}a_{\mathbf{k,\sigma }}$
and $b_{j,s}=\frac{1}{\sqrt{N}}\overset{}{\underset{\mathbf{k}}{\sum }}e^{-i%
\mathbf{kR_{j}}}b_{\mathbf{k,\sigma }}$ where $j=(n,m)$, the Hamiltonian $%
H_{0}$ becomes%
\begin{equation}
H_{0}=-t\underset{\mathbf{k},s}{\overset{}{\sum }}\phi _{\mathbf{k}}a_{%
\mathbf{k},s}^{\dag }b_{\mathbf{k},s}+h.c.+\underset{\mathbf{k},s}{\overset{}%
{\sum }}(\frac{i\lambda _{SO}}{3\sqrt{3}}s\xi _{\mathbf{k}}-elE_{z})(a_{%
\mathbf{k},s}^{\dag }a_{\mathbf{k},s}-b_{\mathbf{k},s}^{\dag }b_{\mathbf{k}%
,s})  \label{sq2}
\end{equation}%
where $\phi _{\mathbf{k}}=\overset{3}{\underset{i=1}{\sum }}e^{i\mathbf{%
k\cdot \delta _{i}}}e^{ik_{z}(h_{A}-h_{B})}=\overset{3}{\underset{i=1}{\sum }%
}e^{i\mathbf{k\cdot \delta _{i}}}e^{ik_{z}2l}$ and $\xi _{\mathbf{k}}=%
\overset{6}{\underset{i=1}{\sum }}e^{i\mathbf{k\cdot n_{i}}}$, where $%
\mathbf{\delta }_{1}=\frac{a}{2}(1,\sqrt{3},0)$, $\mathbf{\delta }_{2}=\frac{%
a}{2}(1,-\sqrt{3},0)$ and $\mathbf{\delta }_{3}=a(1,0,0)$ are the next
nearest neighbor vectors, whereas $\mathbf{n}_{1}=-\mathbf{n_{2}=a_{1}}$, $%
\mathbf{n}_{3}=-\mathbf{n_{4}=a_{2}}$ and $\mathbf{n_{5}=-\mathbf{%
n_{6}=a_{1}-a_{2}}}$ are the six next-nearest neighbor hopping sites (see
figure \ref{siligeom}) that connect identical sublattice sites. Notice that $%
\phi _{\mathbf{k}}~$contains the contribution of the buckled structure in
the $z$ direction given by the factor $e^{ik_{z}(h_{A}-h_{B})}$, where $%
h_{A/B}$ are the sublattice $A$ and $B$ heights with respect to the middle
of the buckling, which obey $h_{A}-h_{B}=2l\,$ (see figure \ref{siligeom}%
)and $k_{z}$ is the wave-vector in the $z$ direction, in contrast to $\xi _{%
\mathbf{k}}$ that does not depend on $l$ because next-nearest neighbors
belong to the same sublattice, 
Ab-initio calculations have shown that there are two most stable sites in
which transition metals can be adsorbed in two-dimensional systems:\ the
center of the hexagon and the bridge between two atoms \cite{A}. In
silicene, the adsorbed atoms preserve the buckled structure, although small
distortions in the geometry near the adsorbed atoms appear changing the
local buckling. This warping of the silicene sheet can alter the distance
between the adatom and the neighboring silicon atoms. The transition metal
atoms most likely hybridize at the hollow site via $s$, $d$ or $f~$orbitals 
\cite{le}. For simplicity we will consider an impurity atom adsorbed in the
top site ($A$ sublattice), with a height $h$ with respect to the $A$ silicon
(see figure \ref{siligeom}) and neglect small distortions of the buckled
structure. In turn, the orbital symmetry that sits on top is not
particularly important and we will consider only an $s$ orbital. Considering
that the adatom is fixed in a position $\mathbf{R}_{0}$ and hybridizes with
the sublattice $A$ with strength $V$, the hybridization Hamiltonian can be
written as%
\begin{equation}
H_{V}=V\overset{}{\underset{s}{\sum }}a_{0,s}^{\dag }(\mathbf{\delta }%
_{0A}^{\prime })f_{s}+h.c.  \label{s2}
\end{equation}%
where $\mathbf{\delta }_{0A}^{\prime }=(h-l)e_{z}$ where $h-l$ is the
distance between the impurity and the $A$ silicon atom and $f_{s}$
annihilates an electron in the magnetic impurity. In the momentum
representation, this last Hamiltonian can be written as%
\begin{equation}
H_{V}=\overset{}{\underset{\mathbf{k},s}{\sum }}Ve^{ik_{z}(h-l)}a_{\mathbf{k}%
,s}^{\dag }f_{s}+h.c.  \label{s2q}
\end{equation}%
Finally, in order to consider the interaction between electrons in the
impurity, we can add the Hamiltonian%
\begin{equation}
H_{F}=\left[ \epsilon _{0}-(1+r)elE_{z}\right] \overset{}{\underset{s}{\sum }%
}n_{s}+Un_{\uparrow }n_{\downarrow }  \label{s2qq}
\end{equation}%
where $\epsilon _{0}$ contains the single electron energy at the impurity
atom, $r=l_{I}/l$, where $l_{I}$ is the distance between the magnetic
impurity and the $A$ sublattice, $n_{s}=$ $f_{s}^{\dag }f_{s}$ is the
occupation number operator for the impurity with spin $s$ and $elE_{z}$ is
the staggered potential. For simplicity we are not considering the
redistribution of charges due to the electric field \cite{nigam}. The
Hubbard parameter $U$ characterizes the strength of the electron
correlations in the inner shell states of the impurity. By adopting the mean
field approximation (\cite{ander}), the Hamiltonian $H_{F}$ can be
decomposed in a constant term and the electronic correlations at the
impurities $Un_{\uparrow }n_{\downarrow }\sim U\overset{}{\underset{s}{\sum }%
}\left\langle n_{s}\right\rangle n_{s}-U\left\langle n_{\uparrow
}\right\rangle \left\langle n_{\downarrow }\right\rangle $, such that the
Hamiltonian of the impurity can be rewritten as $H_{F}=\overset{}{\underset{s%
}{\sum }}\epsilon _{s}n_{s}$, where $\epsilon _{s}=\epsilon +U\left\langle
n_{-s}\right\rangle $, and $\epsilon =\epsilon _{0}-(1+r)elE_{z}$ is the
effective on-site energy of the impurity and the remaining term $%
-U\left\langle n_{\uparrow }\right\rangle \left\langle n_{\downarrow
}\right\rangle $ can be dropped. Then, by considering eq.(\ref{sq2}), eq.(%
\ref{s2q}) and eq.(\ref{s2qq}) the Hamiltonian in compact form can be
written in matrix form in the basis $(\Psi _{A,\uparrow },\Psi _{B,\uparrow
},\Psi _{A,\downarrow },\Psi _{B,\downarrow },\Psi _{I,\uparrow },\Psi
_{I,\downarrow })$ as%
\begin{gather}
H=\underset{\mathbf{k,}s}{\overset{}{\sum }}\left( 
\begin{array}{cccccc}
a_{\mathbf{k},\uparrow }^{\dag } & b_{\mathbf{k},\uparrow }^{\dag } & a_{%
\mathbf{k},\downarrow }^{\dag } & b_{\mathbf{k},\downarrow }^{\dag } & 
f_{\uparrow }^{\dag } & f_{\downarrow }^{\dag }%
\end{array}%
\right) \times   \label{s22} \\
\left( 
\begin{array}{cccccc}
-\Delta _{\mathbf{k}\uparrow } & \phi _{\mathbf{k}}^{\ast } & 0 & 0 & V & 0
\\ 
\phi _{\mathbf{k}} & \Delta _{\mathbf{k}\uparrow } & 0 & 0 & 0 & 0 \\ 
0 & 0 & -\Delta _{\mathbf{k}\downarrow } & \phi _{\mathbf{k}}^{\ast } & 0 & V
\\ 
0 & 0 & \phi _{\mathbf{k}} & \Delta _{\mathbf{k}\downarrow } & 0 & 0 \\ 
V & 0 & 0 & 0 & \epsilon +U\left\langle n_{\downarrow }\right\rangle  & 0 \\ 
0 & 0 & V & 0 & 0 & \epsilon +U\left\langle n_{\uparrow }\right\rangle 
\end{array}%
\right) \left( 
\begin{array}{c}
a_{\mathbf{k},\uparrow } \\ 
b_{\mathbf{k},\uparrow } \\ 
a_{\mathbf{k},\downarrow } \\ 
b_{\mathbf{k},\downarrow } \\ 
f_{\uparrow } \\ 
f_{\downarrow }%
\end{array}%
\right)   \notag
\end{gather}%
where $\Delta _{\mathbf{k}s}=\frac{i\lambda _{SO}}{3\sqrt{3}}s\xi _{\mathbf{k%
}}-elE_{z}$\ and $\phi _{\mathbf{k}}=t\overset{3}{\underset{i=1}{\sum }}e^{i%
\mathbf{k\cdot \delta _{i}}}$. 
\begin{figure}[tbp]
\centering\includegraphics[width=85mm,height=50mm]{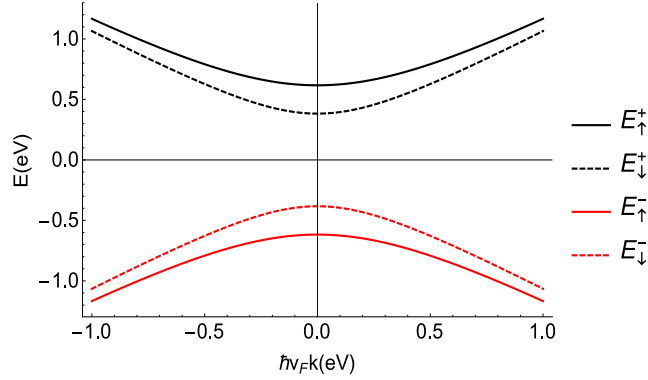}
\caption{Dispersion relation in the long-wavelength approximation for
silicene for each spin, where $elE_{z}=0.5$ eV and~$\protect\lambda =0.039$
eV, ($\pm $) for conduction (valence) band. }
\label{silienergy}
\end{figure}
The local density of states $\rho _{s}(\omega )$ at the impurity can be
obtained as $\rho _{s}(\omega )=-\frac{1}{\pi }\Im {g}_{s}(\omega )$, where $%
g_{s}=\left\langle f_{s}\right\vert G\left\vert f_{s}\right\rangle $ is the
Green function element for each spin $s$ at the impurity level. By solving $%
G=(zI-H)^{-1}$ in the $a_{\mathbf{k},s}$, $b_{\mathbf{k},s}$ and $f_{s}$
basis, a coupled algebraic system is obtained, where the matrix element $%
g_{s}=\left\langle f_{s}\right\vert G\left\vert f_{s}\right\rangle $ reads%
\begin{equation}
g_{s}=\frac{1}{z-\epsilon _{s}-\Sigma _{s}}  \label{s6}
\end{equation}%
where $z=\omega +i0^{+}$ and $\Sigma _{s}$ is the self-energy 
\begin{equation}
\Sigma _{s}=\overset{}{\underset{\mathbf{k,}\alpha =\pm 1}{\sum }}\frac{%
\sigma _{\mathbf{k}\alpha s}}{z-\alpha \epsilon _{\mathbf{k}s}}  \label{s7}
\end{equation}%
where%
\begin{equation}
\sigma _{\mathbf{k}\alpha s}=\frac{V^{2}}{2}\left( 1+\frac{\alpha \Delta _{s}%
}{\epsilon _{\mathbf{k}s}}\right)   \label{s8}
\end{equation}%
where $\Delta _{s}=elE_{z}-s\lambda _{so}$ and where%
\begin{equation}
\epsilon _{\mathbf{k}s}=\sqrt{(E_{z}-s\lambda _{so})^{2}+\hslash
v_{F}^{2}k^{2}}  \label{s8.1}
\end{equation}%
is the low energy dispersion relation of electrons in silicene with
spin-orbit coupling, obtained expanding the numerator and denominator of $%
\xi _{\mathbf{k}}$ and $\phi _{\mathbf{k}}$ around the $K$ point in the
Brillouin zone. 
\begin{figure}[tbp]
\centering\includegraphics[width=160mm,height=120mm]{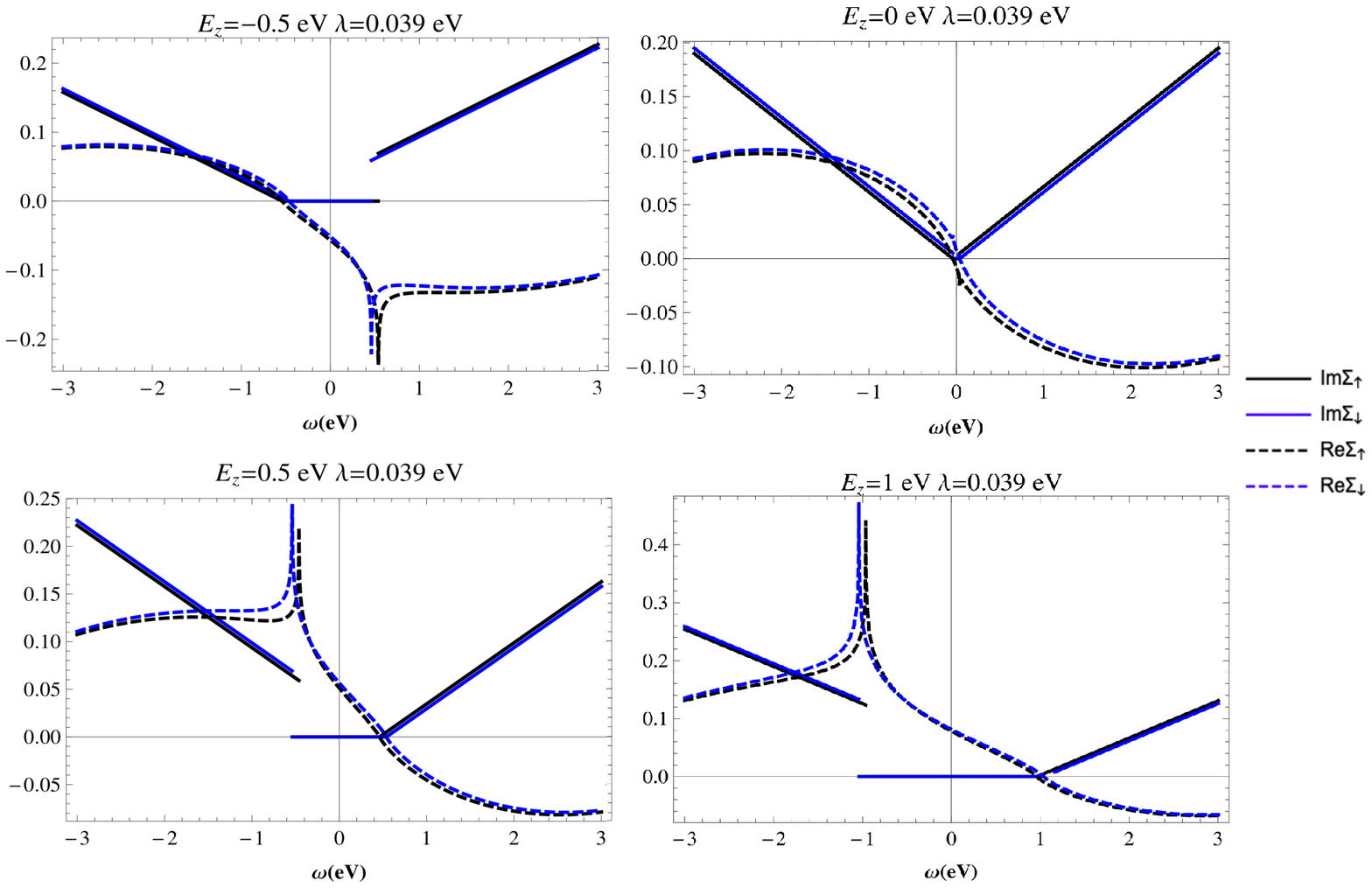}
\caption{Real $\Re \Sigma _{s}$ and imaginary $\Im \Sigma _{s}$ part of the
self-energy for different staggered potential values and where we have used
that $D=7$eV, $V=0.9$ eV, $\protect\epsilon _{0}=0.2$ eV, $U=0$ and $\protect%
\lambda =0.039$ eV and $t=1.6~$eV for silicene. An asymmetric contribution
of $\Sigma _{s}$ on the valence and conduction bands is shown.}
\label{rese}
\end{figure}
From the last equation we can note that there are four bands for $\alpha
,s=\pm 1$ describing electrons ($\alpha =1$) or holes ($\alpha =-1$) with
spin $s$. The bandgap $2\left\vert \Delta _{s}\right\vert \sim 1.5$ meV for $%
elE_{z}0.5$eV turns silicene into a semiconductor, in contrast to graphene,
and the dependence of the gap with the spin is explicit (see figure \ref%
{silienergy}). By computing the imaginary part of the local Green function $%
g_{s}$ at the impurity, the local spin density of states can be obtained as 
\begin{equation}
\rho _{s}(\omega )=-\frac{1}{\pi }\Im {g}_{s}=\frac{\Im \Sigma _{s}}{%
(Z_{s}^{-1}(\omega )\omega -\epsilon _{s})^{2}+\Im ^{2}\Sigma _{s}}
\label{s9}
\end{equation}%
where $Z_{s}^{-1}(\epsilon )=1-\frac{\Re \Sigma _{s}}{\omega }$ is the
quasiparticle residue and $\Re \Sigma _{s}(\Im \Sigma _{s})$ is the real
(imaginary) part of the self-energy which can be written as%
\begin{equation}
\Re \Sigma _{s}=\overset{}{\underset{k,\alpha =\pm 1}{\sum }}\frac{\sigma _{%
\mathbf{k}\alpha s}}{\omega -\alpha \epsilon _{\mathbf{k}s}}\text{ \ \ \ \ \
\ \ }\Im \Sigma _{s}=\pi \overset{}{\underset{k,\alpha =\pm 1}{\sum }}\sigma
_{\mathbf{k}\alpha s}\delta (\omega -\alpha \epsilon _{\mathbf{k}s})
\label{s10}
\end{equation}%
Computing the integral of eq.(\ref{s10}), the real and imaginary part of the
self energy reads 
\begin{gather}
\Re \Sigma _{s}=\frac{V^{2}}{D^{2}}(\Delta _{s}-\omega )\ln \left(
\left\vert \frac{\omega ^{2}-\Delta _{s}^{2}-D^{2}}{\omega ^{2}-\Delta
_{s}^{2}}\right\vert \right)   \label{re1} \\
\Im \Sigma _{s}=\frac{\pi V^{2}}{D^{2}}(\Delta _{s}-\omega )\left[ \theta
(\left\vert \Delta _{s}\right\vert -\omega )-\theta (\left\vert \Delta
_{s}\right\vert +\omega )\right] \theta (D-\left\vert \omega \right\vert ) 
\notag
\end{gather}%
where $D\sim 7$eV is the bandwidth, and where $\theta (\sqrt{\Delta
_{s}^{2}+D^{2}}-\omega )$ and $\theta (\sqrt{\Delta _{s}^{2}+D^{2}}+\omega )$
have been disregarded from the last equation because they only introduce
changes for $\left\vert \omega \right\vert >D$. The last results are a
generalization of \cite{ucho3} for the top site with $\Delta _{s}=0$ in
graphene. In figure (\ref{rese}) the real and imaginary part of the
polarized self-energy are shown for different values of the staggered
potential. In contrast to graphene, $\Im \Sigma _{s}$ and $\Re \Sigma _{s}$
are not symmetric with respect to the Dirac point as it happens for adatoms
on top carbon atoms \cite{bruno}. This is due to the presence of $\Delta _{s}
$ which causes an asymmetry that increases with the external electric field
applied (see figure \ref{rese}). Eq.(\ref{re1}) indicates that the level
broadening scales as $\left\vert \Delta _{s}-\omega \right\vert $ and is
identical to zero for $\left\vert \omega \right\vert <\left\vert \Delta
_{s}\right\vert $. 
\begin{figure}[tbp]
\centering\includegraphics[width=80mm,height=50mm]{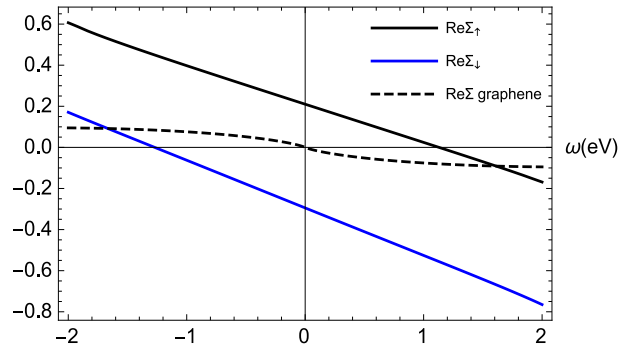}
\caption{Polarized real part $\Re \Sigma _{s}$ of $\Sigma _{s}$ in silicene
compared with $\Re \Sigma $ in graphene near the Dirac point and vanishing
electric field, and where $D=7$eV, $V=0.9$ eV, $\protect\epsilon _{0}=0.2$
eV, $U=0$ and $\protect\lambda =0.039$ eV. At the Dirac point, the
quasiparticle residue in silicene with $E_{z}=0$ is not zero. }
\label{comp}
\end{figure}
The real part of the self-energy shifts the assumed unperturbed energy $%
\omega $ and, in contrast to graphene, it is not identical to zero at the
Dirac point (see figure \ref{comp}). The particle-hole symmetry breaking
occurs in the whole spectrum in contrast with $s$ orbitals for hollow site
adatoms in graphene, where the asymmetry is only evident in the high energy
sector (\cite{bruno}, \cite{romero}).

\section{Results and discussions}

The spin-polarized occupation numbers can be computed using $\rho _{s}$ of
eq.(\ref{s9}) as%
\begin{equation}
n_{s}=\int_{-D}^{\mu }\rho _{s}(\omega )d\omega  \label{r1}
\end{equation}%
where $\mu $ is the Fermi level. In order to obtain unequal spin occupation
numbers at the impurity $n_{\uparrow }\neq n_{\downarrow }$, we must
determine eq.(\ref{s9}), where the polarized density of states $\rho _{s}$
depend on $n_{-s}$. The computation of $n_{\uparrow }$ and $n_{\downarrow }$
implies solving a self-consistent equation for $n_{s}$ as a function of $\mu 
$. The Fermi energy can be tuned experimentally by applying an external
voltage to the sample that adds or subtracts charge carriers, in the form of
electrons or holes (\cite{nair}, \cite{pike}). 
\begin{figure}[h]
\centering\includegraphics[width=130mm,height=70mm]{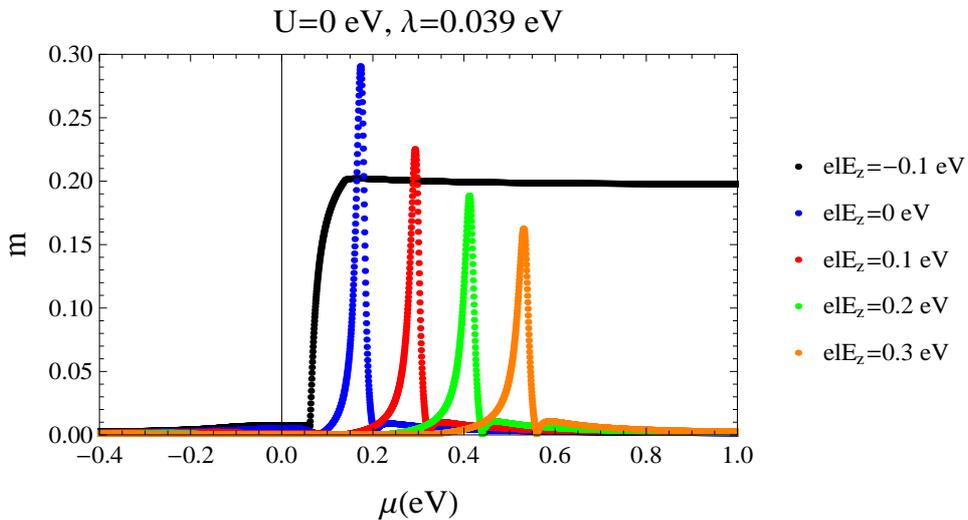}
\caption{Local magnetic moment (in units of $\protect\mu _{B}$) as a
function of $\protect\mu $ for different values of $elE_{z}$ in the $U=0$
limit and where $V=1$ eV and $\protect\epsilon _{0}=0.2$ eV.}
\label{u0}
\end{figure}
Before computing the self-consistency equations for the occupation numbers,
we can study the limits $U=0$ and $U\rightarrow \infty $. In the limit $U=0$%
, local magnetism is possible due to the shift between polarized local
density of states in the impurity created by $\lambda _{so}$. 
\begin{figure}[h]
\centering\includegraphics[width=130mm,height=70mm]{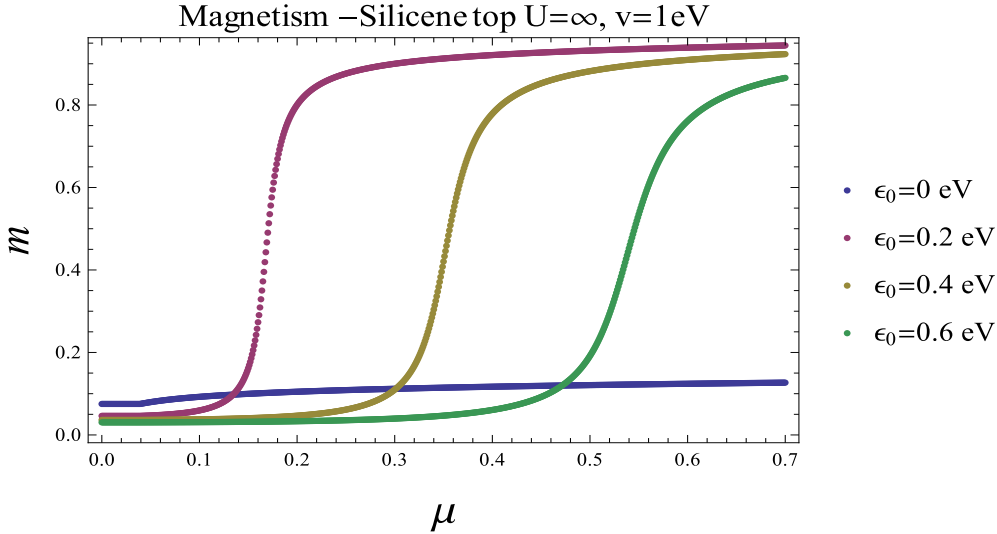}
\caption{Local magnetic moment (in units of $\protect\mu _{B}$) as a
function of $\protect\mu $ for different values of $\protect\epsilon _{0}$
in the $U=\infty $ limit and where $V=1$ eV and $E_{z}=0$. }
\label{uinf}
\end{figure}
Because $U=0$, the occupation numbers $n_{-s}$ do not appear inside $\rho
_{s}$ and eq.(\ref{r1}) can be integrated without difficulty. In figure \ref%
{u0}, the local magnetic moment in units of $\mu _{B}$, where $\mu _{B}$ is
the Bohr magneton, for $U=0$ is shown as a function of $\mu $ for different
values of $elE_{z}$ where the peaks for $elE_{z}>0$ correspond to the shift
of the polarized density of states due to external electric field. On the
other side, in the limit $U=\infty $, one of the quantities $n_{\uparrow }$
or $n_{\downarrow }$ is zero because, in the case $n_{\uparrow }\neq 0$, by
putting a spin-down electron on the adatom implies infinite energy. Then, we
can write without loss of generality that $n_{\downarrow }=0$, then the
local magnetic moment can be computed as $m=n_{\uparrow }(U=\infty )$ (see
figure \ref{uinf}), where the local magnetic moment is shown as a function
of $\mu $ for different values of $elE_{z}$ and tends to $m=1$ for large $%
elE_{z}$ and vanishes for $E_{z}=0$.

\begin{figure}[h]
\centering\includegraphics[width=140mm,height=120mm]{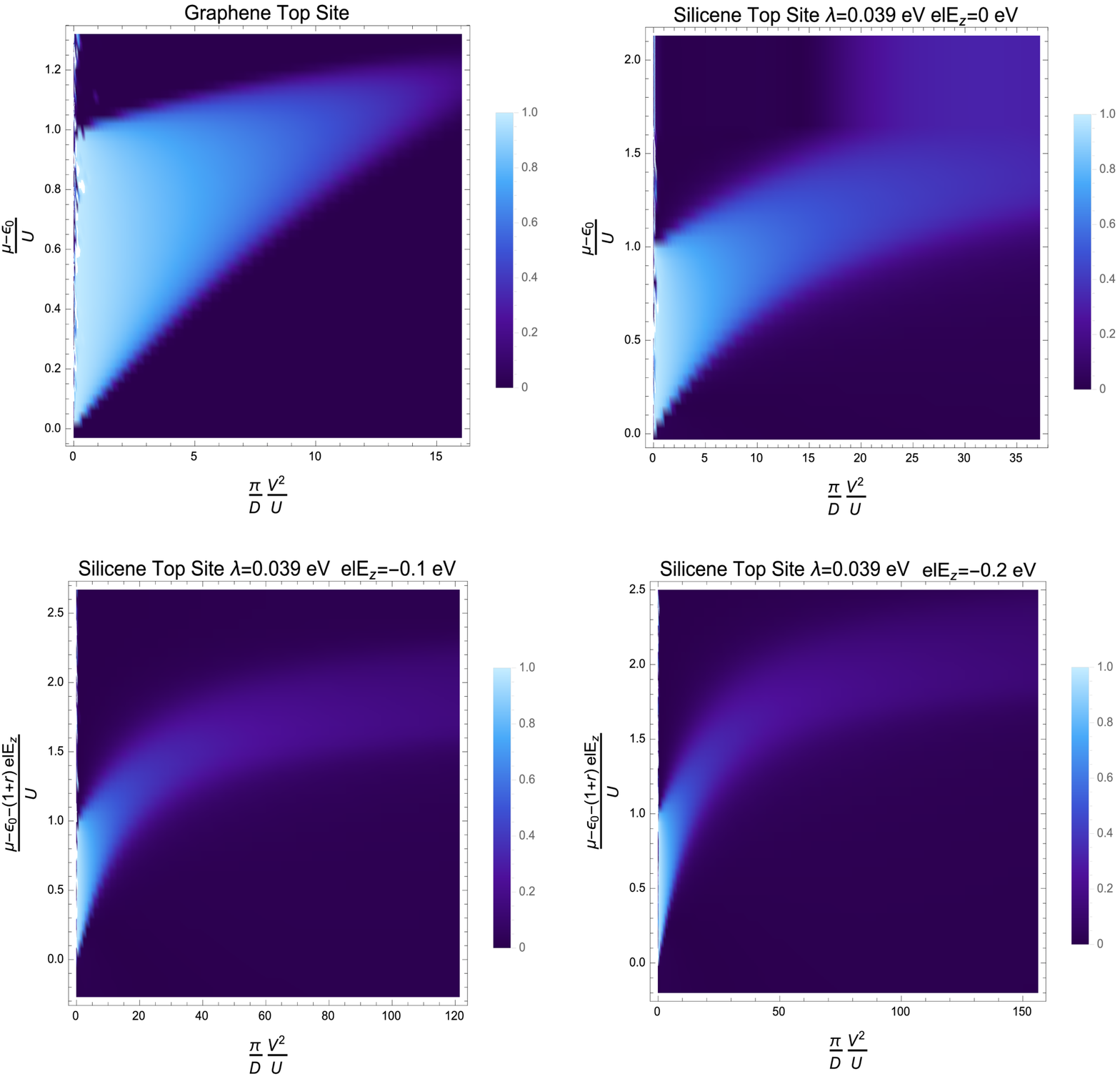}
\caption{Local magnetic moment (in units of $\protect\mu _{B}$) in the
impurity atom in the variables $x=\frac{\protect\pi V^{2}}{UD}$ and $y=(%
\protect\mu -\protect\epsilon _{0})/U$ graphene and silicene with $E_{z}=0$
and $x=\frac{\protect\pi V^{2}}{UD}$ and $y=(\protect\mu -\protect\epsilon %
_{0}-(1+r)elE_{z})/U$, where $\protect\epsilon _{0}=-0.2$ eV, which is below
the Dirac point. }
\label{boundarysilineg1}
\end{figure}

In order to describe local magnetism between these two limits, the
self-consistent equations for $n_{s}$ must be computed by starting with
random values of $n_{\uparrow }$ and $n_{\downarrow }$ and computing $\rho
_{s}(\omega )$ from eq.(\ref{s9}). This local density of states at the
impurity is used through eq.(\ref{s9}) to obtain new values of $n_{\uparrow
} $ and $n_{\downarrow }$ which are reintroduced in ${\rho }_{s}$. 
\begin{figure}[h]
\centering\includegraphics[width=140mm,height=120mm]{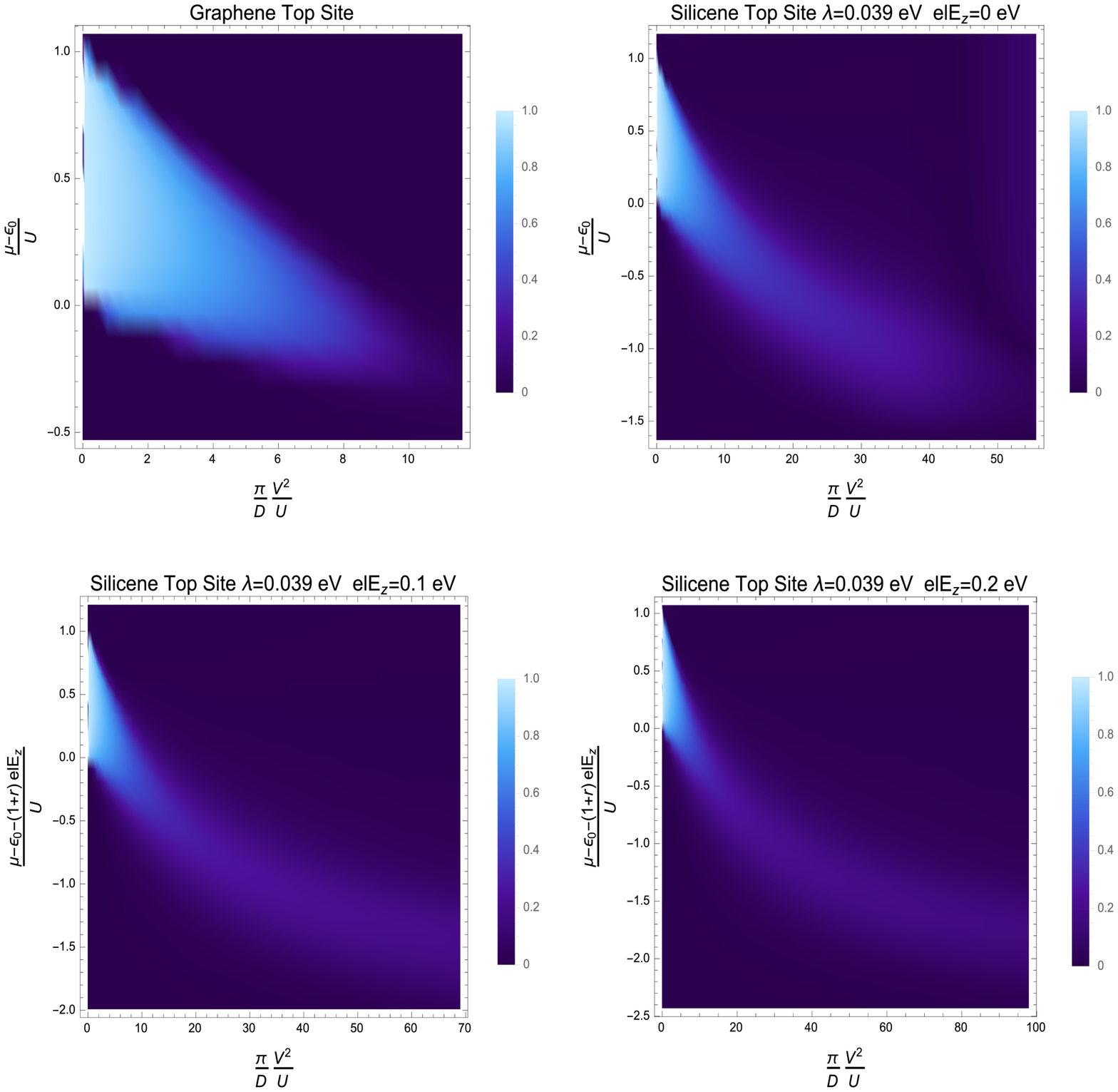}
\caption{Local magnetic moment (in units of $\protect\mu _{B}$) in the
impurity atom in the variables $x=\frac{\protect\pi V^{2}}{UD}$ and $y=(%
\protect\mu -\protect\epsilon _{0})/U$ for graphene and silicene with $%
E_{z}=0$ and $x=\frac{\protect\pi V^{2}}{UD}$ and $y=(\protect\mu -\protect%
\epsilon _{0}-(1+r)elE_{z})/U$, where $\protect\epsilon _{0}=0.2$ eV which
is above the Dirac point. The color bar indicates local magnetic moment in
units of $\protect\mu _{B}$.}
\label{boundarysili}
\end{figure}
The iteration is done until the occupation numbers satisfy the condition $%
\left\vert n_{s}(i+1)-n_{s}(i)\right\vert <10^{-6}$. In figures \ref%
{boundarysili} (for $\epsilon _{0}=0.2$ eV ) and \ref{boundarysilineg1} (for 
$\epsilon _{0}=-0.2$ eV) the magnetic regime of the impurity atom as a
function of $x=\pi V^{2}/UD$ and $y=(\mu -\epsilon _{0}-(1+r)elE_{z})/U$ is
shown for different values of the electric field strength. 
In both figures we can compare the magnetic phases of the impurity atom in
silicene, with and without electric field, with the magnetic phases in
graphene, in terms of the Hamiltonian parameters $x=\frac{\pi V^{2}}{UD}$
and $y=\frac{\mu -\epsilon _{0}-(1+r)elE_{z}}{U}$, where the local magnetic
moment is given in units of $\mu _{B}$ (see the color bar in both figures). 
\begin{figure}[h]
\centering\includegraphics[width=150mm,height=70mm]{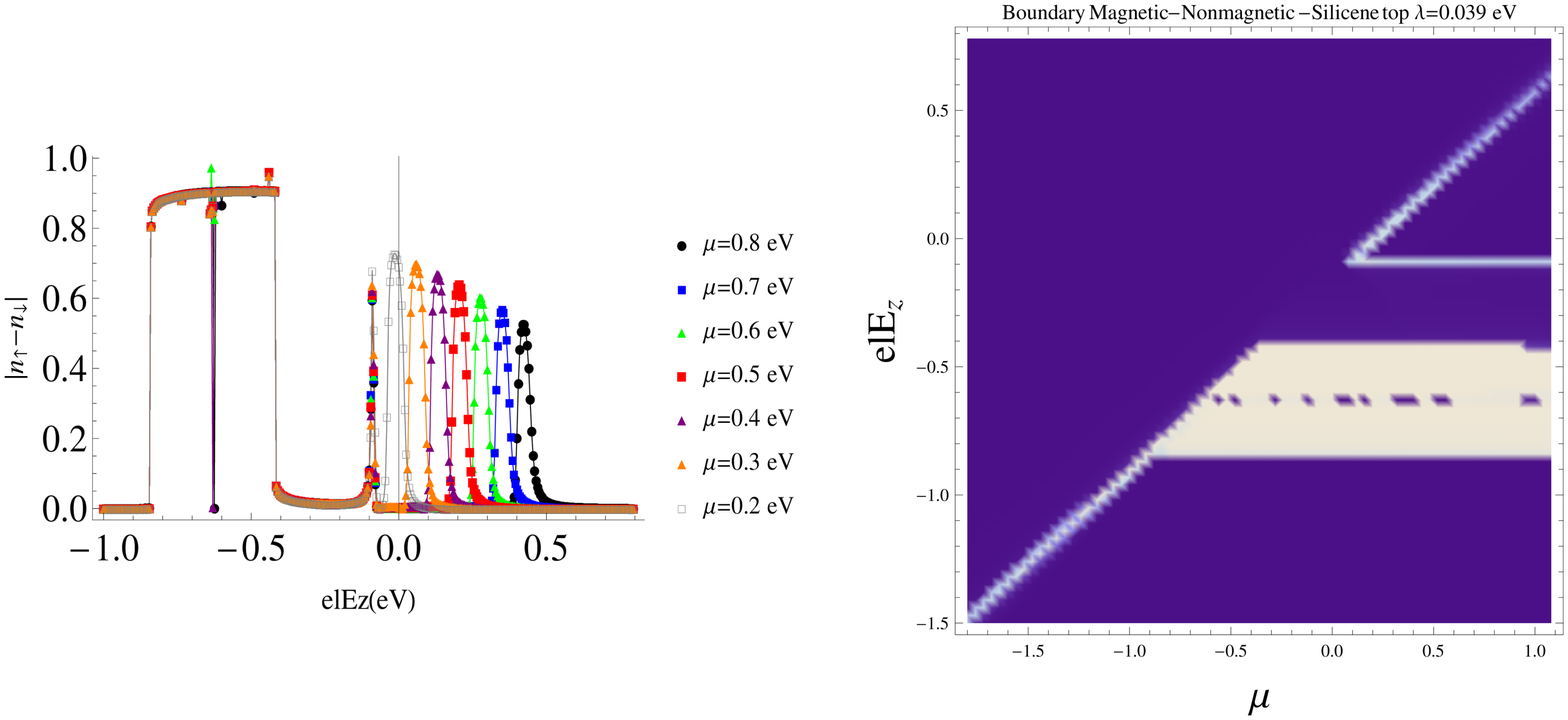}
\caption{Left. Local magnetic moment $\left\vert n_{\uparrow }-n_{\downarrow
}\right\vert $ (in units of $\protect\mu _{B}$) as a function of $elE_{z}$
for different values of $\protect\mu $ where $V=1$eV, $U=0.1$eV, $\protect%
\lambda =0.039$eV, $r=0.3$ and $\protect\epsilon _{0}=-0.2$ eV.
Right.Boundary between magnetic and non-magnetic zone in a $\protect\mu $-$%
elE_{z}$ space for the same set of parameters $V$, $U$, $\protect\lambda $, $%
r$ and $\protect\epsilon _{0}$. The straight line of local magnetism
corresponds to the shifted peaks in the left figure. }
\label{muvsel}
\end{figure}
In both figures, for $x=0$ there is a non-vanishing local magnetic moment
for $\epsilon _{0}<\mu <\epsilon _{0}+U$ without electric field and for $%
\epsilon _{0}+(1+r)elE_{z}<\mu <\epsilon _{0}+U+(1+r)elE_{z}$. This is
expected because the electric field only introduces a shift in the local
energy of the impurity. As in graphene, the boundary in silicene is not
symmetrical at $y=\frac{1}{2}$ and exhibits no particle-hole symmetry around 
$\mu =\epsilon _{0}+(1+r)elE_{z}$ (see \cite{ucho2}) even for a top site
adatom, where the orbital symmetry is irrelevant and the $C_{3v}$ point
group symmetry of the honeycomb sublattice is preserved by the adatom in the
top site. Without electric field, there is a non-vanishing local magnetic
moment when the Fermi level is below $\epsilon _{0}$ for $\epsilon _{0}>0$
as it is shown in figure (\ref{boundarysili}), which is a similar effect
that than found in \cite{jsasili} for hollow site adsorption in silicene. In
turn, the spin-orbit interaction streches the boundary between phases
towards the lower half plane and narrows it when the electric field is
turned on.\footnote{%
The electric field values used are smaller than the critical electric field
at which the honeycomb structure of silicene becomes unstable $elE_{z}\sim
0.59$ eV (see \cite{drum}).} The fact that the boundary between magnetic
phases enlarges for silicene with $\lambda _{so}\neq 0$ and $E_{z}=0$
implies that the suppresion of the local broadening for $\left\vert \omega
\right\vert <\lambda _{so}$ and the unequal spin shift factor $\Delta
_{s}-\omega $ in eq.(\ref{s10}) allows the formation of spin moments for $%
\mu <\epsilon _{0}-U$ and small $U$.

Due to the linear scaling of the broadening with $\omega $ of the impurity
level, magnetism is allowed for Fermi energies $\mu >\epsilon _{0}-\frac{3}{2%
}U$ well below the impurity energy $\epsilon _{0}$ and this behavior is
enhanced when the electric field is turned on.In figure \ref%
{boundarysilineg1}, the same effect is shown for $\epsilon _{0}=-0.2$ eV,
where magnetism can be found when the Fermi level is larger than $\epsilon
_{0}+U$ and the boundary between magnetic phases shrinks in the $y$
direction when the electric field is negative. By increasing the electric
field strength, an asymmetric broadening of the impurity energy level is
lifted by the modification of the imaginary part of the self-energy and in
turn a shift in the impurity peak appears due to effective impurity energy $%
\epsilon _{0}+(1+r)elE_{z}$

In figure (\ref{muvsel}), the local magnetic moment is shown for different
values of $\mu $ as a function of $elE_{z}$, where we have considered $V=1$
eV, $\epsilon _{0}=-0.2$ eV, $\lambda =0.039$eV and $U=0.1$ eV. In both
figures it can be seen that magnetism follows a linear relation between $\mu 
$ and $E_{z}$, caused by the shift in the density of states due to the
effective impurity $\epsilon _{0}+(1+r)elE_{z}$. When $\epsilon
_{0}+(1+r)elE_{z}+Un_{-s}<\mu <Un_{s}$, the ocuppation numbers are not
identical. The boundary of the magnetic phases in this case is controlled by
the broadening of the peak. For $-0.85$eV$<elE_{z}<0.35$eV there is a local
magnetic moment for $\mu >-0.4$ eV. When the impurity peaks in the polarized
density of states of the impurity enters the gap zone, given by $-\left\vert
\Delta _{s}\right\vert <\omega <\left\vert \Delta _{s}\right\vert $, and
when $\mu >-\left\vert \Delta _{s}\right\vert $, the non-vanishing local
magnetic moment is freezed for larger $\mu $. Thus by manipulating the Fermi
level with the applied gate voltage, an adatom interacting with silicene
fullfils the requirement for the formation of a magnetic state due to the
spin-asymmetric anomalous broadening and spin-asymmetric broadening gap for
energies near the Dirac point. In turn, even for small $U$ values \cite%
{graph} and Fermi energies below the effective on-site impurity energy,
magnetism arises in the impurity carried by the itinerant electrons in the
host lattice in contrast with transition metals adatoms, where it is harder
to enhance the local magnetic moment for large $U$ \cite{man}. For low
impurity concentrations, when the local magnetic moments are driven to an
excited state, for example with an external electric field, dynamical
spin-excitations are formed and are carried for long distances \cite{guimar}
which can be utilized in spintronic devices to develop magnetic information
storage with electric gates \cite{cincuanta}. Currently, X-ray magnetic
circular dichroism (XMCD) and inelastic scanning tunneling spectroscopy are
used to identify adatoms with magnetocrystalline anisotropy energy of few
meV deposited on Graphene/SiC showing a paramagnetic behavior, with a
magnetic moment out-of-plane for Co and Fe adatoms \cite{Eel}.

\section{Conclusions}

In this work we have studied the effect of the electric field on the
formation of a local magnetic moment in an impurity adsorbed on a top site
in silicene. By computing the polarized density of states in the impurity
and solving the self-consistent equations for the occupation numbers in the
mean-field approximation, we obtain the boundary of the magnetic phases for
silicene with spin-orbit coupling and different electric field strengths,
considering on-site impurity energies below and above the Dirac point. A
local magnetic moment is formed for Fermi energies below the on-site
impurity energy due to the broadening of the impurity level that scales
linearly in $\left\vert \omega \right\vert $ with a shift due to the
spin-orbit coupling and the external electric field. In turn, a gap in the
broadening for $\left\vert \omega \right\vert <elE_{z}-s\lambda _{so}$
allows the local magnetic moment to be freezed when $\mu $ crosses the gap
zone, even when $E_{z}=0$. By increasing the electric field strength the
boundary between magnetic phases streches allowing a moment formation in
silicene more easily than in graphene. The results obtained can be important
to design spintronic devices, where the local magnetic moment can be
controlled by an electric field application and to manipulate spin waves by
considering different adatoms coverages of silicene subject to external
oscillating electric fields.

\section{Acknowledgment}

This paper was partially supported by grants of CONICET (Argentina National
Research Council) and Universidad Nacional del Sur (UNS) and by ANPCyT
through PICT 1770, and PIP-CONICET Nos. 114-200901-00272 and
114-200901-00068 research grants, as well as by SGCyT-UNS., J. S. A. and P.
J. are members of CONICET., F. E and J. V. are fellow researchers at this
institution.

\section{Author contributions}

All authors contributed equally to all aspects of this work.

\ \ \ \ \ \ \ \ \ \ \ \ \ \ \ \ \ \ \ \ \ \ \ \ \ \ \ \ \ \ \ \ \ \ \ \ \ \
\ \ \ \ \ \ \ \ \ \ \ \ \


\begin{thebibliography}{99}
\bibitem{guz} G. G. Guzm\'{a}n-Verri and L. C. L. Y. Voon, \textit{Phys.
Rev. B} \textbf{76}, 075131 (2007).

\bibitem{hou} M. Houssa, A. Dimoulas, A. Molle, \textit{Jour. Phys. Cond.
Matt.}, \textbf{27} (25), 253002 (2015).

\bibitem{geim} A. Geim, \textit{Science} \textbf{324}, 1530 (2009).

\bibitem{chu} C. V. Nguyen, N. N. Hieu, \textit{Chem. Phys.}, \textbf{468},
9 (2016).

\bibitem{spen} M. J. Spencer, T. Morishita (Eds.), \textit{Silicene},
Springer International Publishing, 2016.

\bibitem{gill} N. Gillgren, D. Wickramaratne, Y. Shi, T. Espiritu, J. Yang,
J. Hu, J. Wei, X. Liu, Z. Mao, K. Watanabe, T. Taniguchi, M. Bockrath, Y.
Barlas, R. K. Lake, C. Ning Lau, \textit{2D Mater.}, \textbf{2 }011001
(2014).

\bibitem{bian} E. Bianco, S. Butler, S. Jiang, O. D. Restrepo, W. Windl, J.
E. Goldberger, \textit{ACS Nano} \textbf{7}, 4413 (2013).

\bibitem{yar11} P. T. T. Le, M. Yarmohammadi, \textit{Chem. Phys.}, \textbf{%
519}, 1-5 (2019).

\bibitem{roldan} R. Rold\'{a}n, L. Chirolli, E. Prada, J. A. Silva-Guill\'{e}%
n, P. San Jose and F. Guinea, \textit{Chem. Soc. Rev.}, \textbf{15} (2017).

\bibitem{yar18} M. Yarmohammadi, \textit{Phys. Rev. B}, \textbf{98}, 155424
(2018).

\bibitem{pad} P. De Padova, C. Quaresima, C. Ottaviani, P. M. Sheverdyaeva,
P. Moras, C. Carbone, D.Topwal, B. Olivieri, A. Kara, H. Oughaddou, B.
Aufray, and G. Le Lay, \textit{Appl. Phys. Lett.}, \textbf{96}, 261905
(2010).

\bibitem{vogt} P. Vogt, P. De Padova, C. Quaresima, J. Avila, E.
Frantzeskakis, M. C. Asensio, A. Resta, B. Ealet, and G. Le Lay, \textit{%
Phys. Rev. Lett.}, \textbf{108}, 155501 (2012).

\bibitem{fleu} A. Fleurence, R. Friedlein, T. Ozaki, H. Kawai, Y. Wang, and
Y. Yamada-Takamura, \textit{Phys. Rev. Lett.}, \textbf{108}, 245501 (2012).

\bibitem{hou2} M. Houssa, E. Scalise, K. Sankaran, G. Pourtois, V. V.
Afanas'ev, A. Stesmans, \textit{Appl. Phys. Lett.}, \textbf{98} (22), 223107
(2011).

\bibitem{wink} R. Winkler and U. Z\"{u}licke, \textit{Phys. Rev. B}, \textbf{%
82}, 245313 (2010).

\bibitem{dz} N. Y. Dzade, K. O. Obodo, S. K. Adjokatse, A. C. Ashu, E.
Amankwah, C. D. Atiso, A. A. Bello, E. Igumbor, S. B. Nzabarinda, J. T.
Obodo, A. O. Ogbuu, O. E. Femi, J. O. Udeigwe, U. V. Waghmare, \textit{%
J.~Phys.: Condens. Matter}, \textbf{22} (37), 375502 (2010).

\bibitem{lebe} S. Lebegue and O. Eriksson, \textit{Phys. Rev. B}, \textbf{79}%
, 115409 (2009).

\bibitem{drum} N. D. Drummond, V. Zlyomi, V. I. Fal'ko, \textit{Phys. Rev. B}%
, \textbf{85} (7), 3702 (2012).

\bibitem{ni} Z. Ni, Q. Liu, K. Tang, J. Zheng, J. Zhou, R. Qin, Z. Gao, D.
Yu, J. Lu, \textit{Nano Letters}, \textbf{12} (1), 113--118 (2012).

\bibitem{liu} C. C. Liu, H. Jiang, Y. Yao, \textit{Phys. Rev. B}, \textbf{84}
(19), 195430 (2011).

\bibitem{yao} Y. Yao, F. Ye, X. L. Qi, S. C. Zhang, Z. Fang, \textit{Phys.
Rev. B}, \textbf{75} (4) 041401 (2007).

\bibitem{liu2} C. C. Liu, W. Feng, Y. Yao, \textit{Phys. Rev. Lett.}, 
\textbf{107} (7), 076802 (2011).

\bibitem{eza} M. Ezawa, \textit{Phys. Rev. B}, \textbf{87} (15) 155415
(2013).

\bibitem{yar4} M. Yarmohammadi and K. Mirabbaszadeh, \textit{Commun. Theor.
Phys.}, \textbf{67}, 5 (2017).

\bibitem{roten} E. Rotenberg, \textit{Graphene Nanoelectronics}, in: H.Raza
(Ed.),Springer-Verlag,Berlin, Heidelberg, 2012.

\bibitem{jsa1} F. Escudero, J. S. Ardenghi, L. Sourrouille, P. Jasen and A.
Juan, \textit{Super. and Micro.}, \textbf{113}, 291-300 (2018).

\bibitem{jsa2} J. S. Ardenghi, P. Bechthold, E. Gonzalez, P. Jasen, A. Juan, 
\textit{Super. and Micro.}, \textbf{72}, 325-335, (2014).

\bibitem{jsa3} J. S. Ardenghi, P. Bechthold, E. Gonzalez, P. Jasen, A., 
\textit{Eur. Phys. J. B}, \textbf{88}: 47 (2015).

\bibitem{jsa4} J. S. Ardenghi, P. Bechthold, P. Jasen, E. Gonzalez, O.
Nagel, \textit{Physica B}, \textbf{427}, 97-105, (2013).

\bibitem{skr} Y. V. Skrypnyk and V. M. Loktev, \textit{Phys. Rev. B} \textbf{%
73}, 241402(R) (2006).

\bibitem{sofo} J. O. Sofo, G. Usaj, P. S. Cornaglia, A. M. Suarez, A. D.
Hernandez-Nieves, and C. A. Balseiro, \textit{Phys. Rev. B} \textbf{85},
115405 (2012).

\bibitem{hew} A. C. Hewson, \textit{The Kondo problem to heavy fermions}
(Cambridge University Press, Cambridge, 1997).

\bibitem{Ren} J. Ren, H. Guo, J. Pan, Y. Y. Zhang, X. Wu, H.-G. Luo, S. Du,
S. T. Pantelides, and H.-J. Gao, \textit{Nano Lett.} \textbf{14}, 4011
(2014).

\bibitem{A} B. Aufray, A. Kara, S. Vizzini, H. Oughaddou, C. L\'{e}andri, B.
Ealet and G. Le Lay, \textit{Appl. Phys. Lett.} \textbf{96}~183102 (2010).

\bibitem{B} E. Cinquanta, E. Scalise, D. Chiappe, C. Grazianetti, B. van den
Broek, M. Houssa, M. Fanciulli and A. Molle, \textit{J. Phys. Chem. C}, 
\textbf{117}~16719--24 (2013).

\bibitem{C} N. Y. Dzade, K. O. Obodo, S. K. Adjokatse, A. C. Ashu, E.
Amankwah, C. D. Atiso, A. A. Bello, E. Igumbor, S. B. Nzabarinda, J. T.
Obodo, A. O. Ogbuu, O. E. Femi, J. O. Udeigwe and U. V. Waghmare, \textit{%
J.~Phys.: Condens. Matter}, \textbf{22}~375502 (2010).

\bibitem{D} V. Q. Bui, T. T. Pham, H. V. S. Nguyen and H. Le, \textit{J.
Phys. Chem. C}, \textbf{117}~23364--71 (2013).

\bibitem{liuu} Y. Liu, X. Zhou, M. Zhou, M.-Q. Long, G. Zhou, \textit{%
J.Appl. Phys.}, \textbf{116} (24) 244312 (2014).

\bibitem{fede1} F. Escudero, J. S. Ardenghi and P. Jasen, \textit{Jour.
Magn. Magn. Mat.}, \textbf{454}, 131-138 (2018).

\bibitem{fede2} F. Escudero, J. S. Ardenghi and P. Jasen, \textit{J. Phys.
Condens. Matter}, \textbf{30}, 275803 (2018).

\bibitem{ander} P. W. Anderson, \textit{Phys. Rev.}, \textbf{124}, 41 (1964).

\bibitem{jsasili} J. Villarreal, J. S. Ardenghi and P. Jasen, \textit{%
Superlattice. Microst.}, \textbf{130} (285-296) 2019.

\bibitem{manuel} M. Laubach, J. Reuther, R. Thomale, S. Rachel, \textit{%
Phys. Rev. B} \textbf{90}, 165136 (2014).

\bibitem{kane} C. L. Kane and E. J. Mele, \textit{Phys. Rev. Lett.,} \textbf{%
95}, 226801 (2005).

\bibitem{zare} M. Zare, \textit{Phys. Rev. B}, \textbf{100}, 085434 (2019).

\bibitem{le} H. M. Le, T. T Pham, T. S. Dinh, Y. Kawazoe, and D.
Nguyen-Manh, \textit{J.~Phys.: Condens. Matter}, \textbf{28}(13), 135301
(2016).

\bibitem{nigam} S. Nigam, S. K. Gupta, C. Majumder and R. Pandey, \textit{%
Phys. Chem. Chem. Phys.}, \textbf{17}, 11324 (2015).

\bibitem{ucho3} B. Uchoa, L. Yang, S. W. Tsai, N. M. R. Peres, and A. H.
Castro Neto, \textit{Phys. Rev. Lett.} \textbf{103}, 206804\ (2009).

\bibitem{bruno} B. Uchoa, L. Yang, S.-W. Tsai, N. M. R. Peres, A. H. Castro
Neto, \textit{New Journal of Physics} \textbf{16}, 013045 (2014).

\bibitem{romero} M. A. Romero, A. Iglesias-Garcia and E. C. Goldberg, 
\textit{Phys. Rev. B}, \textbf{83} 125411 (2011).

\bibitem{nair} R. R. Nair, I-L. Tsai, M. Sepioni, O. Lehtinen, J. Keinonen,
A. V. Krasheninnikov, A. H. Castro Neto, M. I. Katsnelson, A. K. Geim, and
I. V. Grigorieva, \textit{Nat. Commun.} \textbf{4}, 2010 (2013).

\bibitem{pike} N. A. Pike and D. Stroud, \textit{Phys. Rev. B}, \textbf{89}
115428 (2014).

\bibitem{ucho2} B. Uchoa, V. N. Kotov, N. M. R. Peres, and A. H. Castro
Neto, \textit{Phys. Rev. Lett.} \textbf{101}, 026805 (2008).

\bibitem{graph} S. K. Pati, T. Enoki and C. N. R. Rao, \textit{Graphene and
its fascinating attributes}, World Scientific Publishing), 2011 (chapter 7).

\bibitem{man} M. Manad\'{e}, F. Vi\~{n}es, and F. Illas, \textit{Carbon}, 
\textbf{95}:525 (2015).

\bibitem{guimar} F. S. M. Guimaraes, D. F. Kirwan, A. T. Costa, R. B.Muniz,
D. L. Mills, and M. S. Ferreira, \textit{Phys. Rev. B} \textbf{81}, 153408
(2010).

\bibitem{cincuanta} Li Tao, E. Cinquanta, D. Chiappe, C. Grazianetti, M.
Fanciulli, M. Dubey, A. Molle and D. Akinwande, \textit{Nature Nanotech.} 
\textbf{10}, 227--231 (2015)

\bibitem{Eel} T. Eelbo, M. Wasniowska, P. Thakur, M. Gyamfi, B. Sachs, T. O.
Wehling, S. Forti, U. Starke, C. Tieg, A. I. Lichtenstein, and R.
Wiesendanger, \textit{Phys. Rev. Lett.} \textbf{110},136804 (2013).
\end{thebibliography}
\end{document}